\documentclass{ieeeaccess}
\usepackage{cite}
\usepackage{amsmath,amssymb,amsfonts}
\usepackage{algorithmic}
\usepackage{graphicx}
\usepackage{multirow}
\usepackage{caption,setspace}
\usepackage{arydshln}
\usepackage{subfigure}
\usepackage{booktabs}
\usepackage{color}

\usepackage{textcomp}
\def\BibTeX{{\rm B\kern-.05em{\sc i\kern-.025em b}\kern-.08em
    T\kern-.1667em\lower.7ex\hbox{E}\kern-.125emX}}
\begin{document}
\history{Date of publication xxxx 00, 0000, date of current version xxxx 00, 0000.}
\doi{10.1109/ACCESS.2017.DOI}

\title{A novel multimodal approach for hybrid brain-computer interface}

\author{\uppercase{Zhe Sun}\authorrefmark{1},
\uppercase{Zihao Huang\authorrefmark{1},  Feng Duan*\authorrefmark{1} and Yu Liu\authorrefmark{2}}}
\address[1]{College of Artificial Intelligence, Nankai University, Jinnan, Tianjin, China}
\address[2]{Key Laboratory of Exercise and Health Sciences of Ministry of Education, Shanghai University of Sport, Shanghai, China}

\tfootnote{This work was supported by the National Key R\&D Program of China (No. 2017YFE0129700), the National Natural Science Foundation of China (Key Program) (No. 11932013), the National Natural Science Foundation of China (No. 61673224), and the Tianjin Natural Science Foundation for Distinguished Young Scholars (No. 18JCJQJC46100)}

\markboth
{Author \headeretal: Preparation of Papers for IEEE TRANSACTIONS and JOURNALS}
{Author \headeretal: Preparation of Papers for IEEE TRANSACTIONS and JOURNALS}

\corresp{Corresponding author: Feng Duan (e-mail: duanf@nankai.edu.cn).}

\begin{abstract}
Brain-computer interface (BCI) technologies have been widely used in many areas. In particular, non-invasive technologies such as electroencephalography (EEG) or near-infrared spectroscopy (NIRS) have been used to detect motor imagery, disease, or mental state. It has been already shown in literature that the hybrid of EEG and NIRS has better results than their respective individual signals. The fusion algorithm for EEG and NIRS sources is the key to implement them in real-life applications. In this research, we propose three fusion methods for the hybrid of the EEG and NIRS-based brain-computer interface system: linear fusion, tensor fusion, and $p$th-order polynomial fusion. Firstly, our results prove that the hybrid BCI system is more accurate, as expected. Secondly, the $p$th-order polynomial fusion has the best classification results out of the three methods, and also shows improvements compared with previous studies. For a motion imagery task and a mental arithmetic task, the best detection accuracy in previous papers were 74.20\% and 88.1\%, whereas our accuracy achieved was 77.53\% and 90.19\% . Furthermore, unlike complex artificial neural network methods, our proposed methods are not as computationally demanding.
\end{abstract}


\titlepgskip=-15pt
\maketitle
\section{Introduction}
Brain-computer interface (BCI) is an important tool for the detection of signal patterns in brain activity. It has been used widely in many areas, such as in robotics control, workload detection or brain-disease detection\cite{muhl2014eeg, ang2015randomized, latchoumane2012multiway, blankertz2010berlin, liu2020efficient, mao2019brain, wu2020electroencephalographic}. BCI technology is usually divided into invasive BCI and non-invasive BCI. Invasive BCI requires the sensors to record brain activities from within the skull. Non-invasive BCI, on the other hand, records brain signals using sensors placed on the scalp, and it is undoubtedly a much safer technology and easier to use \cite{waldert2016invasive}. However, signals from non-invasive BCI sensors are usually full of noise, from the subjects' unconscious eye movement or ambient noise, for example. Finding a way to use non-invasive sensors to detect brain signals accurately is still a big challenge\cite{zhang2015sparse}.
\par
Transitional non-invasive BCI systems use either electroencephalography (EEG) sensors, near-infrared spectroscopy (NIRS) sensors or Magnetic Resonance Imaging (MRI) to record brain activities\cite{zhang2019strength, zhang2014frequency, fazli2012enhanced, sitaram2009hemodynamic}. Compared with the MRI equipment, EEG and NIRS equipments are low-cost and smaller in size. Both EEG and NIRS have been implemented in many real time BCI applications\cite{fazli2012enhanced}. 
\par
EEG equipment consists of metal electrodes placed on the scalp to record electrical signals \cite{zheng2018using}. The electrodes record the activity of the surrounding neurons. For the EEG-based BCI system, classical feature extraction methods such as common spatial patterns (CSP)\cite{CSP1998}, power spectrum density (PSD) and auto-regressive modeling (AR)\cite{MAFeature} have been previously proposed to analyze and localize the EEG patterns and activated brain area of motor imagery (MI) and mental arithmetic (MA)\cite{features}. Then, in regards to classifiers, several machine learning classifiers such as k-nearest neighbours (KNN)\cite{KNN}, support vector machine (SVM) \cite{SVM1,SVM2}, and linear discriminant analysis (LDA) \cite{LDA} are used in classifying these proposed EEG features. In recent years, with the wide application of deep learning, many studies have already shown that convolution neural network (CNN) exhibits good performance for EEG processing or NIRS processing \cite{zhang2019novel, trakoolwilaiwan2017convolutional}. CNN and  CNN-based models, such as EEGNet \cite{lawhern2018eegnet} or convolutional neural networks-stacked autoencoders (CNN-SAE)\cite{cnn-sae}, have obtained remarkable MI classification results. Moreover, some researchers have developed an LSTM-based framework \cite{lstm} to extract essential features of time-varying EEG signals to classify motor imagery signals.
\par
NIRS-based BCI systems, generally used to measure the hemodynamic signals from target regions of the brain \cite{nirs_intro}, can also analyze oxy-hemoglobin NIRS (oxy-NIRS) or deoxy-hemoglobin NIRS (deoxy-NIRS) concentrations in order to localize and classify the brain activity \cite{hishimoto2019mitochondrial}. Generally, when a specific brain area becomes more active, energy metabolism increases, leading to an increased oxygen consumption and increased levels of carbon dioxide in the area. Then, oxy-NIRS signals will decrease while deoxy-NIRS signals will increase. For the signal processing or pattern classification of NIRS, transitional algorithms such as SVM or artificial neural networks are used\cite{naseer2015fnirs}.
\par
Instead of using EEG or NIRS mehods individually, using them simultaneously provides a multimodal hybrid BCI. It has been repeatedly proven in literature that multimodal BCI systems have a better accuracy and stability than the BCIs based on a single modality \cite{openaccess2017, shin2018ternary, chiarelli2018deep}.
For instance, Jaeyoung Shin et al. presented an open source benchmark for a hybrid EEG-NIRS fusion BCI system \cite{openaccess2017, fazli2012enhanced} which also uses a linear discriminant analysis classifier for the multimodal data classifications.
\par
Non-linear feature fusion is commonly used in deep learning, and there are many ways to achieve it. 
For instance, in \cite{lin2015bilinear} a bilinear method was used to fuse two feature vectors, which can be thought of as the 2-order tensor product fusion. Even further, a trilinear method was used in \cite{ben2017mutan} to fuse three modality features. In addition, some tensor-based methods have been proposed to solve multimodal fusion problems such as Visual Question Answering: for example, \cite{bai2018deep} proposed a Deep Attention Neural Tensor Network which can discover the joint correlations over images, questions and answers with tensor-based representations; \cite{liu2018efficient} proposed a low-rank multimodal fusion method which performs multimodal fusion using low-rank tensors to improve efficiency; finally, \cite{hou2019deep} proposed a polynomial tensor pooling block for integrating multimodal features by considering high-order moments.
\par
In this study, a novel approach for the hybrid BCI system is proposed. Firstly, CNN was used to get the feature vectors from the EEG, oxy-NIRS, and deoxy-NIRS signals. A single-modal data was calculated to compare with hybrid-system based classification results. Then, three fusion methods were used to process the multimodal feature vectors: linear fusion classification method, tensor fusion classification method and the $p$th-order polynomial fusion classification method.
Jaeyoung Shin et al's benchmark dataset \cite{openaccess2017} was used for comparison purposes to validate the methods proposed in this paper.
\par
The classification results for both the motor imagery (MI) experiment data and the mental arithmetic (MA) experiment data were compared with results from previous works. In both these two different tasks, the $p$th-order polynomial fusion classification shows the best classification performance. In the MI task, the accuracy achieved was 77.53\%, while the accuracy achieved in the MA task was 90.19\%. Furthermore, compared with the deep neural networks method, our algorithm is simple and uses less computing resources. It can be used for real-time hybrid BCI systems.
\par
The rest of the paper is organized as follows: hybrid BCI benchmark dataset and fusion algorithm are introduced in Section II. Experiments and pre-processing, as well as their results, are presented and discussed
in Section III. The discussion and conclusions are given in the
Section IV and Section V respectively.

\section{Materials and Methods}

\subsection{Datasets}
\textbf{Subjects} 29 subjects, consisting of 14 males and 15 females, participated in the data collection procedures. The average age was $28.5\pm 3.7$ (mean$\pm$ standard deviation).

\textbf{Data Acquisition} EEG data was collected by a thirty-channel BrainAmp EEG amplifier (Brain Products GmbH, Gilching, Germany), with a sampling rate of 1000 Hz. These electrodes were placed according to the international 10-5 system. NIRS data was collected by NIRScout (NIRx GmbH, Berlin, Germany) with a sampling rate of 12.5 Hz and containing thirty-six channels. More details about the electrodes' position can be found in \cite{openaccess2017}.

\begin{figure}
  \centering
    \includegraphics[width=0.5\textwidth]{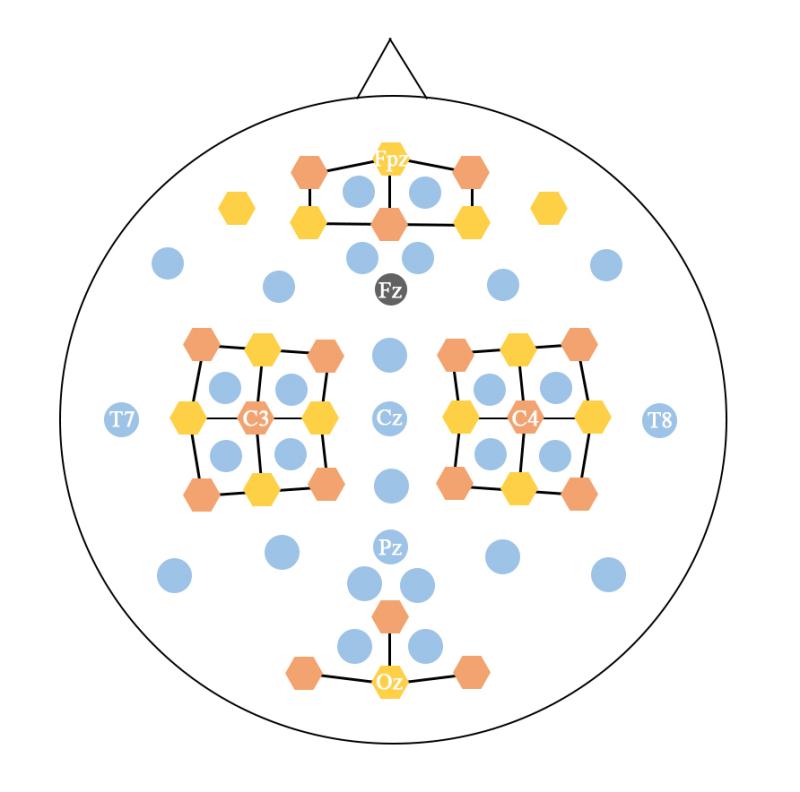}
    \caption{Locations of the EEG electrodes (solid circles), NIRS sources (yellow hexagons) and detectors (orange hexagons).}
    \label{fig:eegandnirs}
\end{figure}

\begin{figure*}[ht]
  \centering
  \includegraphics[width=0.8\linewidth]{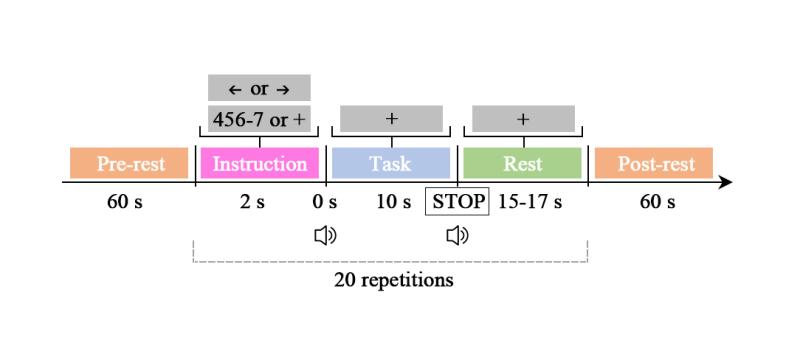}  
  \caption{The flow of experiments.}
  \label{exp_prcess}
\end{figure*}

\begin{figure*}[t]
    \centering
    \includegraphics[width=1\linewidth]{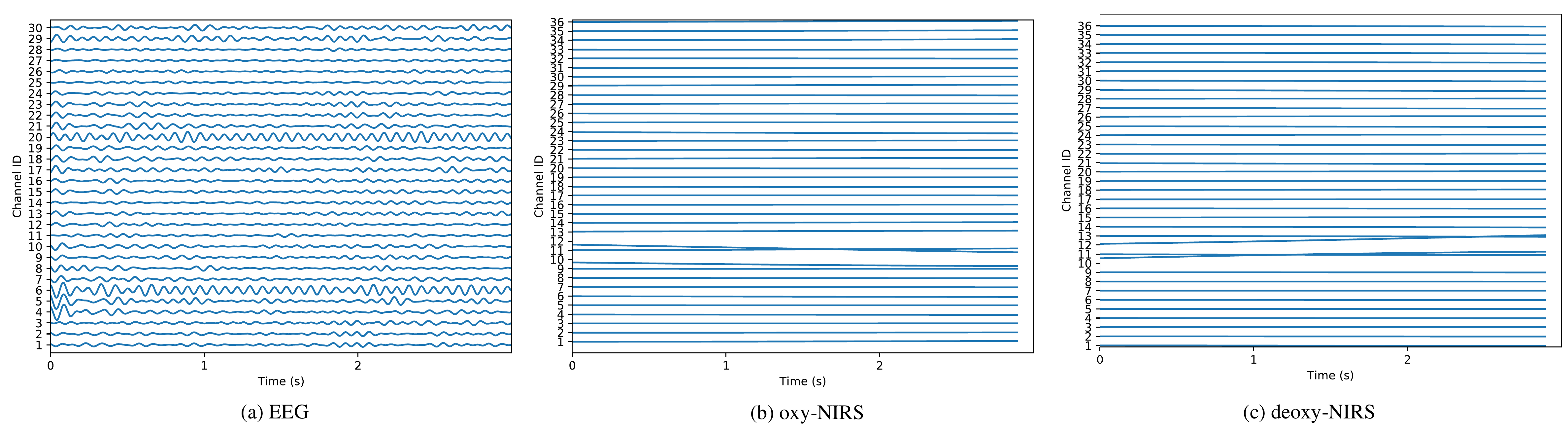}
    \caption{An example of EEG, oxy-NIRS and deoxy-NIRS data. Notice that the NIRS data are of low-frequency and are filtered by a bandpass filter from 0.01 Hz to 0.1 Hz.}
    \label{example}
\end{figure*}
\textbf{Motor imagery (MI) experiment} Subjects were required to conduct kinesthetic MI. Namely, they needed to imagine opening or closing their hand while they were grabbing a ball, so that the MI is actual MI rather than visual. Each trial consisted of three parts: instruction (2s), task (10s) and rest (15-17s). Each session consisted of a pre-rest (60s) , 20 repetitions of the aforementioned trial, and a post-rest (60s). The MI experiment dataset contains the data from 3 sessions.
\par
\textbf{Mental arithmetic (MA) experiment} Subjects were required to conduct subtraction in the form of a three-digit number minus a one-digit number (e.g., 384-8). As in the MI experiment, the MA experiment dataset also contains data on 3 sessions where each session consists of 20 repetitions of the trial. Subjects were given an initial subtraction calculation, and completed this subtraction as their first trial of their total 20. For the remaining 19, the new subtraction was the result from the last calculation minus the initial one-digit number, which thus remained constant throughout the trials.
\par
\textbf{Data pre-processing} The EEG data was first re-referenced by using common average reference and filtered with the 4th-order Chebyshev type 2 filter with a bandpass of 0.5-50 Hz. Then, independent component analysis (ICA)-based EOG rejection was used to remove artifacts. After that, the EEG data was downsampled to 200 Hz. For each trial, 35s of data was extracted, containing a segment of the last rest (10s), the task (10s) and the final rest (15s). Then, a 3s time window with 1s step was employed to collect data. Eventually, 33 segments were obtained from each trial. Each segment has a shape of $30\times600$ (channels$\times$times).  For the NIRS data, the concentration changes of oxy- and deoxy-hemoglobin (oxy-NIRS and deoxy-NIRS) were first calculated by using the modified Beer-Lambert law. The oxy-NIRS and dexoy-NIRS were then filtered by a 6th-order zero-phase Butterworth filter with a bandpass of 0.01-0.1 Hz. After that, a 3s time window like the EEG was used to segment the NIRS data. After this process, a segment ends up with a shape of $36\times30$. Thus, for both the MI and MA experiments, each subject completed 60 trials, and each trial has 33 segments. Fig.~\ref{example} shows the segments of one of the subjects for three modal data from 0s to 3s.
\par
\subsection{Multimodal Classification Method}
In this section, we will first show how the classification model is built by using single modal data. Following this, two common methods to fuse features in deep learning are introduced to allow for comparison with our method and a better understanding of it. After that, our $p$th-order polynomial fusion method is introduced. In the end, in order to tackle the unacceptably large amount of parameters, tensor decomposition is used.


\textbf{Notation} Focusing on the tri-modal task, assume $x^1$, $x^2$ and $x^3$ denote the EEG, oxy-hemoglobin NIRS (oxy-NIRS) and deoxy-hemoglobin NIRS (deoxy-NIRS), respectively. Also assume that $z^1$, $z^2$ and $z^3$ denote their respective feature vectors obtained from three CNNs. To express the formula concisely, Einstein notation \cite{harrison2016numeric} is applied to describe multiplication between tensors or between a tensor and a vector. In particular, if we assume $x_i$ to denote a vector and $\mathcal{W}_{ijk}$ to denote a 3rd-order tensor, their product can be written as $y_{jk}=x_{i}\mathcal{W}_{ijk}$, which means $y_{jk}=\sum_{i}^{I}x_i\mathcal{W}_{ijk}$. Also, when we concatenate $z^1$, $z^2$ and $z^3$, the result is then written as $z^{1,2,3}$. Given a vector $x_{i_1}$, 
its first copy is written as $x_{i_2}$, and its $(N-1)$th copy is $x_{i_N}$. 

\textbf{Single modal classification}
We have data from three different modalities, and each has a shape of channels$\times$times.
Specifically, these are EEG, oxy-NIRS and deoxy-NIRS, with shapes $30\times600$, $36\times30$ and $36\times30$, respectively. For the single-modal classification, we conduct 1D-CNN and then each convolutional layer is followed by a batch-norm layer and a ReLU layer \cite{schirrmeister2017deep}. Since oxy-NIRS and deoxy-NIRS have same shape, we use the same CNN structure for them. Table \ref{EEGnet} and \ref{NIRSnet} show the CNNs used on EEG and NIRS in detail. Notice that the feature vectors $z^1$, $z^2$ and $z^3$ are the results of ``AvgPool1d''.
\par
\begin{figure}
    \centering
    \includegraphics[width=1\linewidth]{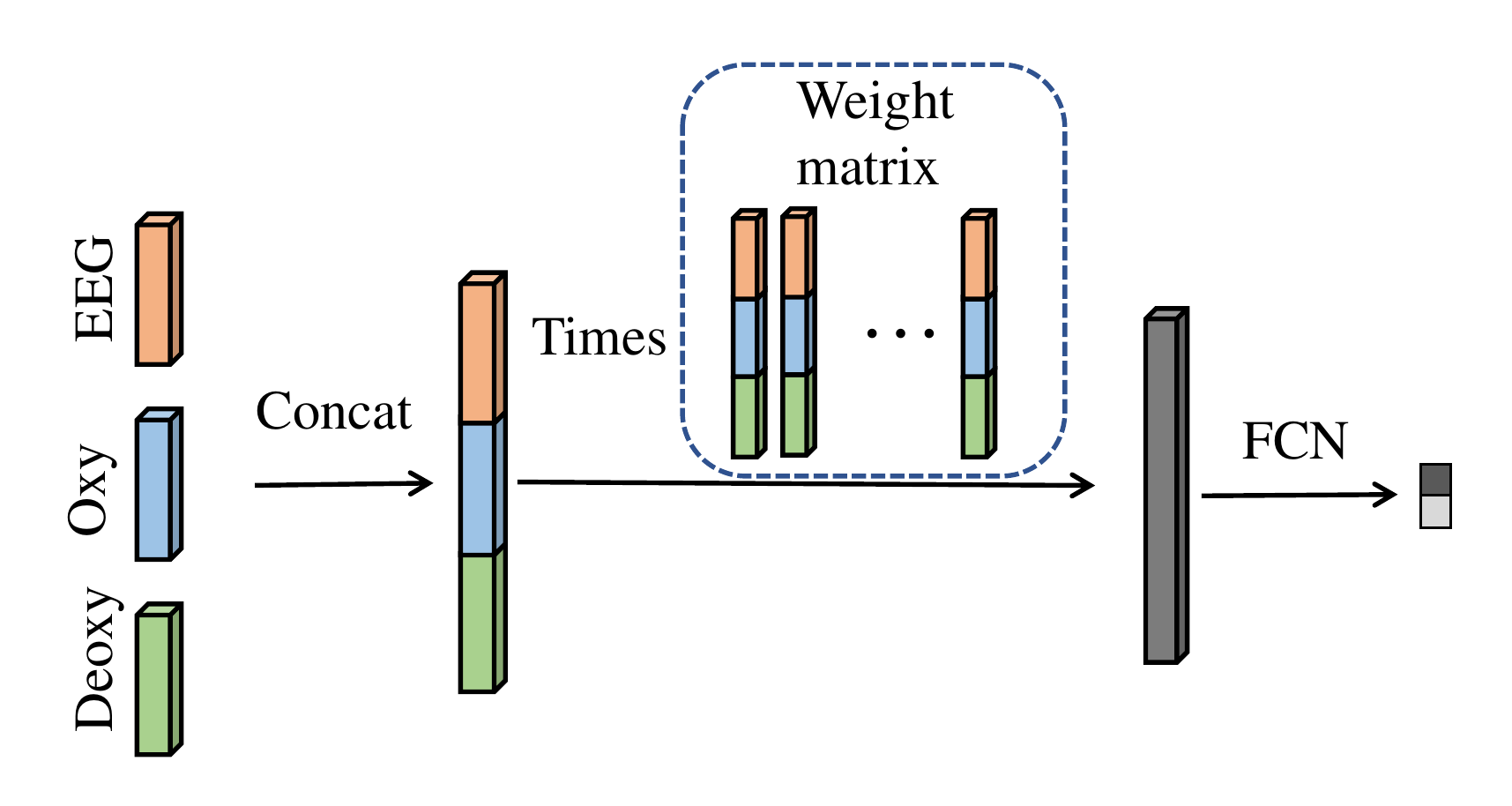}
    \caption{Diagram of the linear fusion (LF) method. Fused data is classified using a fully connected neural network (FCN).}
    \label{LF_demo}
\end{figure}
\begin{figure}
    \centering
    \includegraphics[width=1\linewidth]{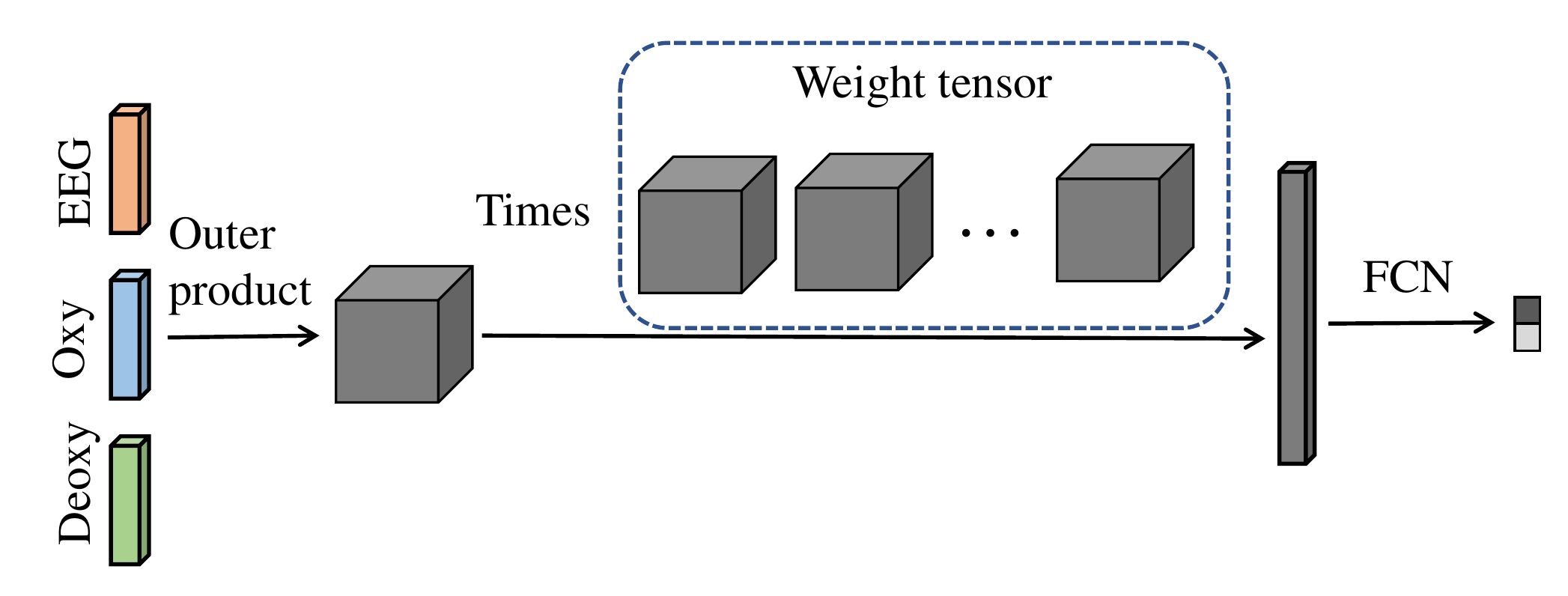}
    \caption{Diagram of a tensor fusion method. Fused data is classified using a fully connected neural network (FCN).}
    \label{TF_demo}
\end{figure}
\begin{figure*}
    \centering
    \includegraphics[width=1.0\linewidth]{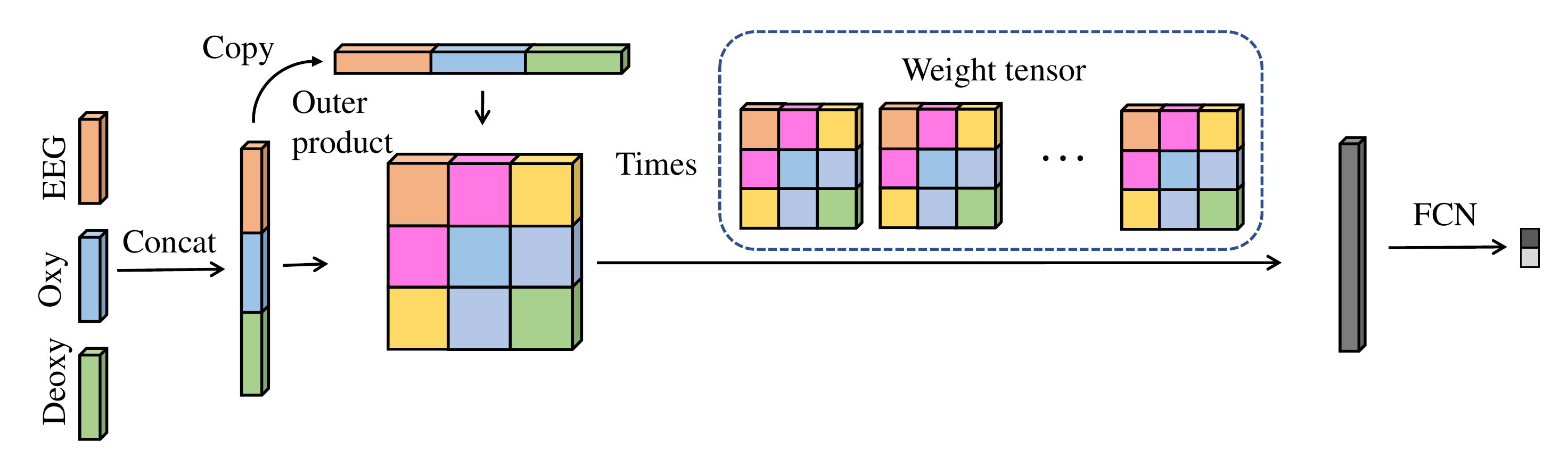}
    \caption{Diagram of the $p$th-order polynomial fusion method. Fused data is classified using a fully connected neural network (FCN).}
    \label{PF_demo}
\end{figure*}

Considering the triple modal fusion problem, there are linear and multilinear methods. Fig. \ref{LF_demo}, \ref{TF_demo}, and \ref{PF_demo} show three ways to fuse multimodal data.
\par

\textbf{Linear fusion (LF) classification} It is commonly used to simply concatenate all feature vectors obtained from the output of ``AvgPool1d''. Thus, the linear fusion product can be written as
\begin{equation}
    y_o=z^{1,2,3}_{i}W_{i,o},
    \label{LF}
\end{equation}
where $W_{i,o}$ is a fusion matrix, and $y_{o}$ is the fused feature vector with a length of $o$. $y_{o}$ is then put in a fully connected network (FCN) in order to generate a classification result. Fig. \ref{LF_demo} shows the flow of LF classification.

\textbf{Tensor fusion (TF) classification}
For the tensor fusion, the outer product is usually introduced. It is a natural way to obtain a feature containing the interaction amongst multiple feature vectors. In our task, the outer product result can be written as 
\begin{equation}
    \mathcal{Z}_{abc}=z^{1}_{a}z^{2}_{b}z^{3}_{c},
\end{equation}
where $\mathcal{Z}_{abc}$ is a 3rd-order feature tensor, and $a,b,c$ are the lengths of three modal feature vectors. After that, a 4th-order weight tensor $\mathcal{W}_{abco}$ is used to obtain the fused feature vector $y_o$:
\begin{equation}
    y_o=\mathcal{Z}_{abc}\mathcal{W}_{abco}.
    \label{TF}
\end{equation}
Then, $L2$ normalization is applied to $y_o$. Finally, FNC is used for classification. This operation is the same as the one shown above in the LF classification section. Fig. \ref{TF_demo} shows the flowchart of TF classification.

\textbf{$p$th-order polynomial fusion ($p$th-PF) classification}
Tensor fusion considers the interaction between multiple feature vectors. However, the interactions within each feature vector or between two of the feature vectors are not present in the fusion. To tackle this problem, polynomial fusion is introduced. As in with linear fusion, we firstly obtain $z^{1,2,3}$ by concatenating all feature vectors. Then, $p-1$ copies of $z^{1,2,3}$ are made and the outer product is calculated:
\begin{equation}
    \mathcal{Z}^{1,2,3}_{i_{1}i_{2}...i_{p}}=z^{1,2,3}_{i_1}z^{1,2,3}_{i_2}\cdots z^{1,2,3}_{i_p}.
\end{equation}
Also, as in with tensor fusion, a $(p+1)$th-order weight tensor $\mathcal{W}_{i_{1}i_{2}...i_{p}o}$ is employed for fusion:
\begin{equation}
    y_o=\mathcal{Z}^{1,2,3}_{i_{1}i_{2}...i_{p}}\mathcal{W}_{i_{1}i_{2}...i_{p}o}.
    \label{PF}
\end{equation}
The subsequent operations are the same as shown in the TF classification section. Fig. \ref{PF_demo} shows the flowchart of PF classification.

\textbf{Rethink three fusion methods}
As it turns out, the fusion weight matrix/tensor in these fusion methods can be ``decomposed''. It should be noticed that the ``decomposition'' here is just to decompose the aforementioned formulas in order to make them easier to understand. For LF, Eq.~\ref{LF} can be written as:
\begin{equation}
    y_o=z^{1}_{a}W^{1}_{ao}+z^{2}_{b}W^{2}_{bo}+z^{3}_{c}W^{3}_{co},
\end{equation}
where
\begin{equation}
        W_{io}=\begin{bmatrix}
W^{1}_{ao}\\ 
W^{2}_{bo}\\ 
W^{3}_{co}
\end{bmatrix}.
\end{equation}
\par
We can see that LF actually only considers the single modal feature and simply adds them together.
For TF, the $\mathcal{W}_{abco}$ in Eq.~\ref{TF} cannot be ``decomposed'' since three modal vectors are entangled together. This may raise the problem of overfitting in the model, because the noise in the sample may be amplified during feature entanglement.
For $p$th-PF, LF is a special case of $p$th-PF when $p=1$. To be concise with the formula, if we let $p=2$, then Eq.~\ref{PF} can be written as:
\begin{equation}
\begin{aligned}
y_o=&\mathcal{Z}^{1,1}_{a_1a_2}\mathcal{W}^{1,1}_{a_1a_2o}+
\mathcal{Z}^{1,2}_{a_1b_2}\mathcal{W}^{1,2}_{a_1b_2o}+
\mathcal{Z}^{1,3}_{a_1c_2}\mathcal{W}^{1,3}_{a_1c_2o}+\\
&\mathcal{Z}^{2,1}_{b_1a_2}\mathcal{W}^{2,1}_{b_1a_2o}+
\mathcal{Z}^{2,2}_{b_1b_2}\mathcal{W}^{2,2}_{b_1b_2o}+
\mathcal{Z}^{2,3}_{b_1c_2}\mathcal{W}^{2,3}_{b_1c_2o}+\\
&\mathcal{Z}^{3,1}_{c_1a_2}\mathcal{W}^{3,1}_{c_1a_2o}+
\mathcal{Z}^{3,2}_{c_1b_2}\mathcal{W}^{3,2}_{c_1b_2o}+
\mathcal{Z}^{3,3}_{c_1c_2}\mathcal{W}^{3,3}_{c_1c_2o},
\end{aligned}
\end{equation}
where
\begin{equation}
\mathcal{Z}^{1,2,3}_{i_1i_2}=\begin{bmatrix}
 \mathcal{Z}^{1,1}_{a_1a_2}& \mathcal{Z}^{1,2}_{a_1b_2} & \mathcal{Z}^{1,3}_{a_1c_2}\\ 
 \mathcal{Z}^{2,1}_{b_1a_2}& \mathcal{Z}^{2,2}_{b_1b_2} & \mathcal{Z}^{2,3}_{b_1c_2}\\ 
 \mathcal{Z}^{3,1}_{c_1a_2}& \mathcal{Z}^{3,2}_{c_1b_2} & \mathcal{Z}^{3,3}_{c_1c_2}
\end{bmatrix},
\end{equation}
and
\begin{equation}
    \mathcal{W}^{1,2,3}_{i_1i_2o}=\begin{bmatrix}
 \mathcal{W}^{1,1}_{a_1a_2o}& \mathcal{W}^{1,2}_{a_1b_2o} & \mathcal{W}^{1,3}_{a_1c_2o}\\ 
 \mathcal{W}^{2,1}_{b_1a_2o}& \mathcal{W}^{2,2}_{b_1b_2o} & \mathcal{W}^{2,3}_{b_1c_2o}\\ 
 \mathcal{W}^{3,1}_{c_1a_2o}& \mathcal{W}^{3,2}_{c_1b_2o} & \mathcal{W}^{3,3}_{c_1c_2o}
\end{bmatrix}.
\end{equation}
We can therefore see that all interactions between modes are taken into account. Compared with TF, this kind of entanglement may also introduce more noise, but at the same time it also generates more fruitful features. It can be observed, from the experimental results, that this operation has indeed a positive effect.

\textbf{Dimension reduction by tensor decomposition} In deep learning, the usually large amount of parameters can lead to difficult training or overfitting \cite{shi2017deep, zhang2015sparse}. The number of parameters required by TF increases exponentially along with the increase of the mode, and the number of parameters required by PF increases exponentially along with the increase of the order. In order to tackle this issue, canonical/polyadic (CP) \cite{kolda2009tensor} decomposition is used. Its main idea is based on using several core tensors to represent the original weight tensor. Given a weight tensor $\mathcal{W}_{i_1i_2...i_po}$, this tensor can be represented by:
\begin{equation}
    \mathcal{W}_{i_1i_2...i_po}=\mathcal{W}_{i_1ro}\mathcal{W}_{i_2ro}\cdots \mathcal{W}_{i_pro}\mathcal{W}_{r},
\end{equation}
where $r$ is called \textit{rank}, and is used to control the size of all core tensors. With the increase of $r$, the core tensors can better approximate the weight tensor (i.e. increasing the ability of representation). Thus, Eq.~\ref{PF} can be written as:
\begin{equation}
\begin{aligned}
    y_o=&\mathcal{Z}^{1,2,3}_{i_{1}i_{2}...i_{p}}\mathcal{W}_{i_1ro}\mathcal{W}_{i_2ro}\cdots\mathcal{W}_{i_pro}\mathcal{W}_{r}\\
=&(z^{1,2,3}_{i_1}\mathcal{W}_{i_1ro})(z^{1,2,3}_{i_2}\mathcal{W}_{i_2ro})\cdots(z^{1,2,3}_{i_p}\mathcal{W}_{i_pro})\mathcal{W}_{r}
\end{aligned}
\end{equation}
Considering that all of $z^{1,2,3}_{i_p}$ are the same, we can further reduce the parameters by assuming symmetric structure of the core tensors, namely, $\mathcal{W}_{i_mro}=\mathcal{W}_{i_nro}, m\neq n$.

\textbf{Complexity analysis} Assume that three feature vectors have lengths of $A$, $B$ and $C$, and the fusion vector has a length of $O$. Let CP \textit{rank} be $R$. Then, the computational and storage complexity for LF, TF and $p$th-order PF will be $\mathcal{O}((A+B+C)O)$, $\mathcal{O}(ABCO)$ and $\mathcal{O}((ABC)^pO)$, respectively. By conducting CP decomposition, the computational and storage complexity for TF and $p$th-order PF will be $\mathcal{O}((A+B+C)RO)$ and $\mathcal{O}(p(A+B+C)RO)$. By assuming symmetric structure of the core tensors, the complexity for $p$th-order PF is reduced to $\mathcal{O}((A+B+C)RO)$. We can thus see that CP decomposition significantly reduces the amount of parameters, and also that symmetric structure can further reduce parameters on $p$th-order PF.

\begin{table*}[ht!]
\centering
\caption{1D-CNN neural network structure to extract the EEG signal features.}
\label{EEGnet}
\begin{tabular}{|c|c|c|c|c|c|c|}
\hline
Layer & Input (C$\times$T) & Operation & Filter size & Stride & Padding & Output \\ \hline
\multirow{2}{*}{1} & 30$\times$600 & Conv1D & 9 & 4 & 0 & 60$\times$148 \\ \cline{2-7} 
 & 60$\times$148 & BatchNorm+ReLU &  &  &  & 60$\times$148 \\ \hline
\multirow{2}{*}{2} & 60$\times$148 & Conv1D & 3 & 1 & 0 & 60$\times$146 \\ \cline{2-7} 
 & 60$\times$146 & BatchNorm+ReLU &  &  &  & 60$\times$146 \\ \hline
\multirow{2}{*}{3} & 60$\times$146 & Conv1D & 3 & 1 & 0 & 60$\times$144 \\ \cline{2-7} 
 & 60$\times$144 & BatchNorm+ReLU &  &  &  & 60$\times$144 \\ \hline
\multirow{2}{*}{4} & 60$\times$144 & Conv1D & 9 & 4 & 0 & 120$\times$32 \\ \cline{2-7} 
 & 120$\times$32 & BatchNorm+ReLU &  &  &  & 120$\times$32 \\ \hline
\multirow{2}{*}{5} & 120$\times$32 & Conv1D & 3 & 1 & 0 & 120$\times$30 \\ \cline{2-7} 
 & 120$\times$30 & BatchNorm+ReLU &  &  &  & 120$\times$30 \\ \hline
\multirow{2}{*}{6} & 120$\times$30 & Conv1D & 3 & 1 & 0 & 120$\times$28 \\ \cline{2-7} 
 & 120$\times$28 & BatchNorm+ReLU &  &  &  & 120$\times$28 \\ \hline
7 & 120$\times$28 & AvgPool1D &  &  &  & 120 \\ \hline
8 & 120 & Linear & 120 &  &  & 60 \\ \hline
9 & 60 & ReLU &  &  &  & 60 \\ \hline
10 & 60 & Linear & 60 &  &  & 2 \\ \hline
11 & 2 & Softmax &  &  &  & 2 \\ \hline
\end{tabular}
\end{table*}

\begin{table*}[ht]
\centering
\caption{1D-CNN neural network structur to extract the NIRS signal features.}
\label{NIRSnet}
\begin{tabular}{|c|c|c|c|c|c|c|}
\hline
Layer & Input (C$\times$T) & Operation & Filter size & Stride & Padding & Output \\ \hline
\multirow{2}{*}{1} & 36$\times$30 & Conv1D & 9 & 4 & 0 & 72$\times$13 \\ \cline{2-7} 
 & 72$\times$13 & BatchNorm+ReLU &  &  &  & 72$\times$13 \\ \hline
\multirow{2}{*}{2} & 72$\times$13 & Conv1D & 3 & 1 & 0 & 72$\times$11 \\ \cline{2-7} 
 & 72$\times$11 & BatchNorm+ReLU &  &  &  & 72$\times$11 \\ \hline
\multirow{2}{*}{3} & 72$\times$11 & Conv1D & 3 & 1 & 0 & 72$\times$9 \\ \cline{2-7} 
 & 72$\times$9 & BatchNorm+ReLU &  &  &  & 72$\times$9 \\ \hline
\multirow{2}{*}{4} & 72$\times$9 & Conv1D & 9 & 4 & 0 & 144$\times$7 \\ \cline{2-7} 
 & 144$\times$7 & BatchNorm+ReLU &  &  &  & 144$\times$7 \\ \hline
\multirow{2}{*}{5} & 144$\times$7 & Conv1D & 3 & 1 & 0 & 144$\times$5 \\ \cline{2-7} 
 & 144$\times$5 & BatchNorm+ReLU &  &  &  & 144$\times$5 \\ \hline
\multirow{2}{*}{6} & 144$\times$5 & Conv1D & 3 & 1 & 0 & 144$\times$3 \\ \cline{2-7} 
 & 144$\times$3 & BatchNorm+ReLU &  &  &  & 144$\times$3 \\ \hline
7 & 144$\times$3 & AvgPool1D &  &  &  & 144 \\ \hline
8 & 144 & Linear & 144 &  &  & 72 \\ \hline
9 & 72 & ReLU &  &  &  & 72 \\ \hline
10 & 72 & Linear & 72 &  &  & 2 \\ \hline
11 & 2 & Softmax &  &  &  & 2 \\ \hline
\end{tabular}
\end{table*}
\par

\section{Experiments}
\subsection{Network configuration}
For the single-modal model, the network details are shown in Table~\ref{EEGnet} and Table \ref{NIRSnet}. For the triple-modal model, the CNN model for extracting feature vectors is the same as the one before ``Avgpool1d'' in the Table~\ref{EEGnet} and Table \ref{NIRSnet}. The fused feature vector has a length of 128. TF employs CP decomposition, and $p$th-order PF employs CP decomposition as well as symmetric structure. After that, a FCN with one layer is employed for classification.
\par
In the training phase, we applied cross-entropy as the loss function, and the Adam optimizer to train the model, where the learning rate was set to $0.001$ and other parameters were default. All models were trained by 300 epochs with a mini-batch size of 16.
\par
For the triple-modal data, we have 29 subjects, with each subject carrying out 60 trials. A trial lasts 35 seconds, where resting goes from -10s to 0s, the task goes from 0s to 10s, followed by another resting period from 10s to 25s. We only used 3s data as the input for the model. For each subject, we conducted a 5-fold cross validation and took the average. Finally, all the subjects' results were averaged.
\par

\subsection{Results}
We firstly compared the classification accuracy results with the best results from the following published paper \cite{openaccess2017}. Our methods show significant improvement.

\begin{figure*}
    \centering
    \includegraphics[width=0.8\linewidth]{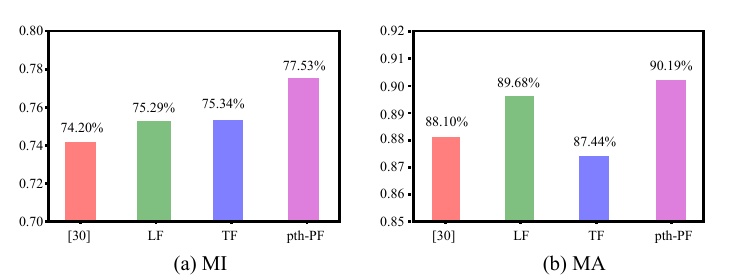}
    \caption{Classification accuracy for the MI (figure a) and MA (figure b) tasks. Comparison between our methods and the ones in \cite{openaccess2017}. MI: Motion Imagery task; MA: Mental arithmetic task.}
    \label{compare}
\end{figure*}

\begin{figure*}
\centering
  \includegraphics[width=1.0\linewidth]{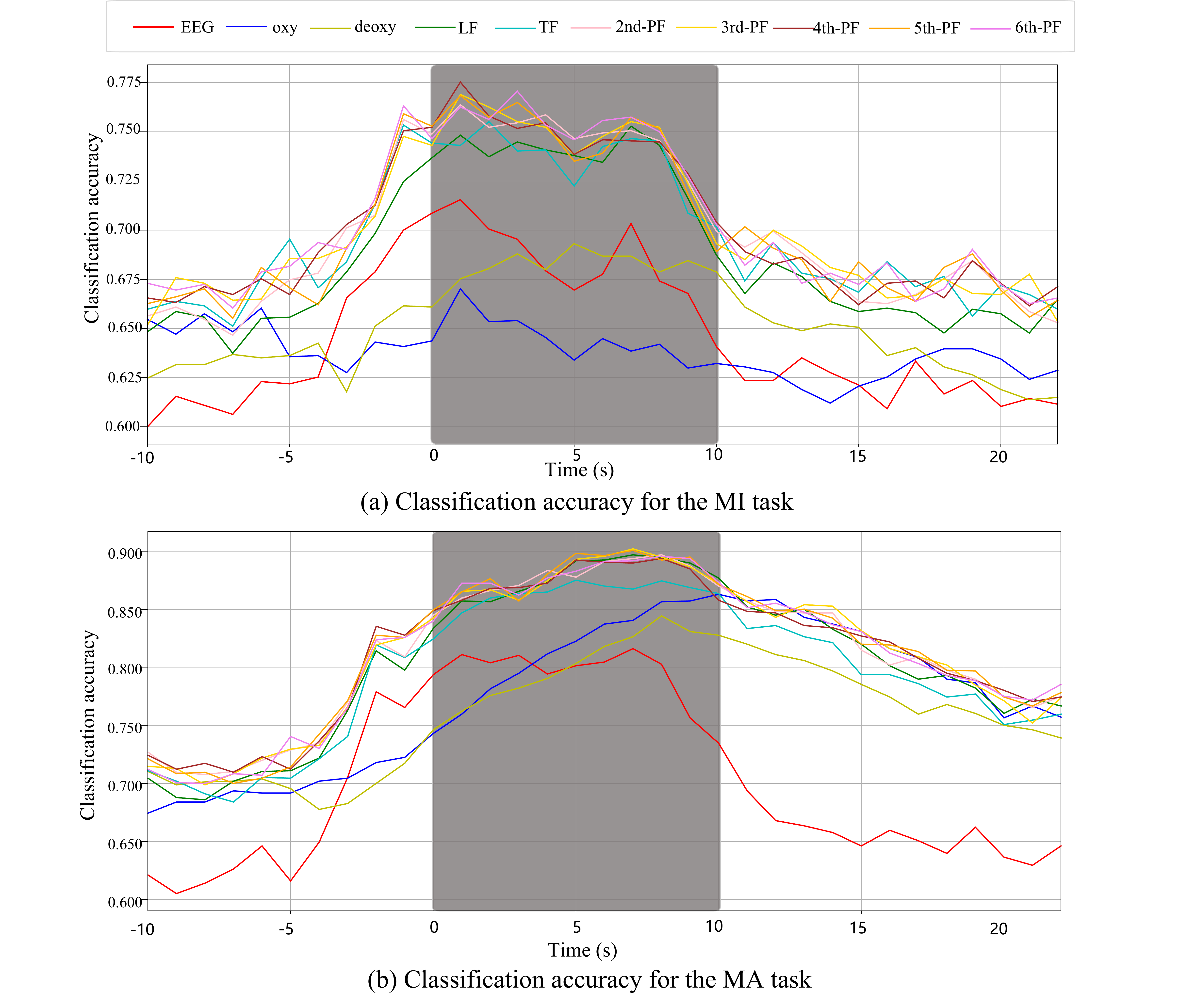}  
  \caption{(a): Classification accuracy for the motion imagery (MI) task. (b): Classification accuracy for mental arithmetic (MA) task.} 
  \label{group}
\end{figure*}
The results of the classification accuracies of our methods are displayed in Fig. \ref{group}, where x-axis indicates the left edge of the moving time window and the y-axis shows the accuracy. We can see that all the tri-modal fusion models perform better than single-modal models. Furthermore, in both tasks, PF achieves better results than LF and TF.
\par
In the MI task, the best results for EEG, oxy-NIRS, deoxy-NIRS, LF, TF and PF are obtained by the time window 1 (71.55\%), 1 (67.01\%), 5 (69.31\%), 7 (75.29\%), -1 (75.34\%) and 1 (77.53\%)(5th-order), respectively. 
\par
In the MA task, these respective time windows for optimal results are 7 (81.60\%), 10 (86.28\%), 8 (84.42\%), 7 (89.68\%), 8 (87.44\%) and 8 (90.19\%)(3rd-order), respectively. The best results for LF, TF, PF and \cite{openaccess2017} are shown in Fig.~\ref{compare}.


\section{Discussion}
\subsection{Combining EEG and NIRS signals}
In recent years, a large number of new technologies have been developed for the analysis of brain activity from single-modal BCI equipment. New algorithms are always being developed over time that improve the reliability of the current BCI system. However, a single-model BCI system still has many limitations. For instance, a multimodal BCI has better anti-noise capabilities than a single-model BCI \cite{shin2018improvement, khan2014multi, wallois2012usefulness, wallois2012usefulness}. Multimodality implies a multi-view, and thus a different representation for the same brain activity patterns\cite{fazli2013brain}. It helps the algorithm to detect specific patterns from the signals. Many previous studies are in line with our results having also indicated the improvement by fusion of multi-signals\cite{ahn2017multi}.
\par
We selected a NIRS and EEG-based hybrid BCI fusion system because both of these two systems are real-time and low-cost\cite{shin2018ternary, pei2020brainkilter}. Additionally, compared with MRI, these two technologies do not require a medical license. We believe EEG and NIRS-based hybrid BCI methods can and will be used as a critical tool in many applications.

\subsection{Hybrid BCI signal processing based on tensor fusion}
According to recent literature, deep learning methods or artificial neural networks have been applied for the detection of patterns in brain activity. Several artificial neural networks have also been applied to this dataset. However, almost all of the neural networks result in an over-fitting problem since the BCI dataset includes small samples. Although tuning the neural structures or parameters could slightly improve the accuracy, the method's stability leaves much to be desired \cite{zhang2019novel, appriou2020modern}.
\par
A tensor is a higher-order array that represents signals from different types of sensors\cite{lahat2015multimodal, acar2017tensor, hunyadi2017tensor}. For example, EEG data or NIRS data can be represented by time × frequency
× electrode, and functional MRI data can be represented by voxels × scans ×
subjects \cite{beckmann2005tensorial, miwakeichi2004decomposing, acar2007multiway}. In addition, tensor decomposition has been widely used to capture brain signal patterns.
In this work, the classification accuracy of our proposed methods is much better than neural networks, and our system is also very robust.

\section{Conclusions}
In this paper we have focused on using tensor fusion methods to construct a hybrid BCI system. The hybrid BCI system includes simultaneous EEG signals and NIRS signals. It also includes two tasks: a MI task and a MA task. We used a shallow CNN neural network to detect the EEG and NIRS feature vectors. Then, we used LF, TF and $p$th-order PF fusion methods to integrate the feature vectors. The results show that the $p$th-order PF fusion method has the best classification accuracy out of the three fusion methods and the single-modal classification.
Furthermore, when comparing the performance of our method with published literature, $p$th-order shows better results. We believe our method could be useful for hybrid BCI systems.


\section*{Acknowledgements}
The authors would like to thank Dr. Kai Zhang for the fruitful advice and Pau Solé-Vilaró for the English review.

\subsection{References}
\bibliographystyle{ieeetr}
\bibliography{sun}

\begin{IEEEbiography}[{\includegraphics[width=1in,height=1.25in,clip,keepaspectratio]{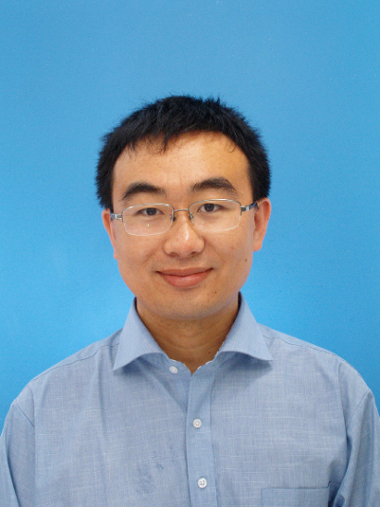}}]{Zhe Sun} received the Ph.D. degree from
Yokohama City University, in 2017. He joined
RIKEN, in 2015, as a Research Support Assistant.
He has been a Research Scientist with RIKEN,
since 2017. He is currently a Research Scientist
with the Computational Engineering Applications
Unit, R\&D Group, Head Office for Information
Systems and Cybersecurity, National Institute of
RIKEN, Japan. From 2014 to 2017, his research
topics were the development of spiking neuron
model, and spiking neural network to understand end elucidate brain functions. His current research interests include large scale brain simulation and
neuromorphic engineering.
\end{IEEEbiography}

\begin{IEEEbiography}[{\includegraphics[width=1in,height=1.25in,clip,keepaspectratio]{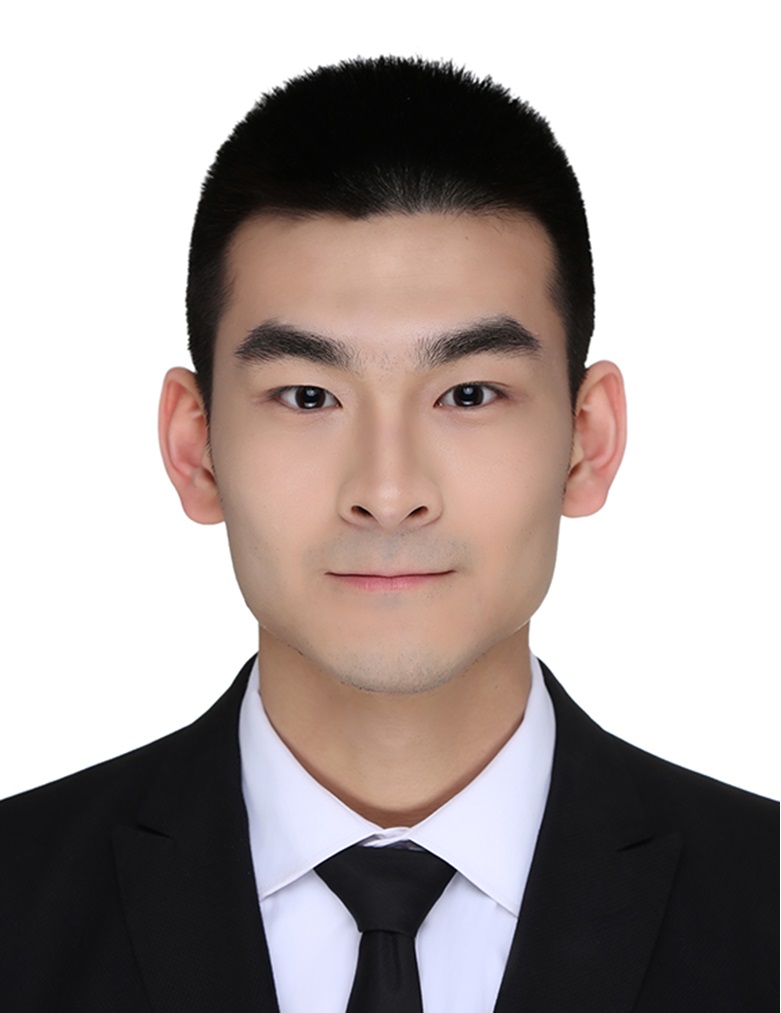}}]{Zihao Huang} received the B.E. degree in automation from Nankai University, Tianjin, China, in 2018, where he is currently pursuing the master's degree with the College of Artificial Intelligence. His current research interests include EEG signal processing and robot research.
\end{IEEEbiography}

\begin{IEEEbiography}[{\includegraphics[width=1in,height=1.25in,clip,keepaspectratio]{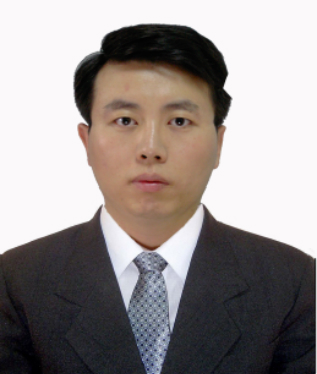}}]{Feng Duan} received the B.E. and M.E. degrees in mechanical engineering from Tianjin University, China, in 2002 and 2004, respectively. He received the M.S. and Ph.D. degrees in precision engineering from the University of Tokyo, Japan in 2006 and 2009, respectively. Currently, he is a professor at Nankai University, P. R. China. His research interests include cellular manufacture systems, rehabilitation robots, and brain machine interfaces.
\end{IEEEbiography}

\begin{IEEEbiography}[{\includegraphics[width=1in,height=1.25in,clip,keepaspectratio]{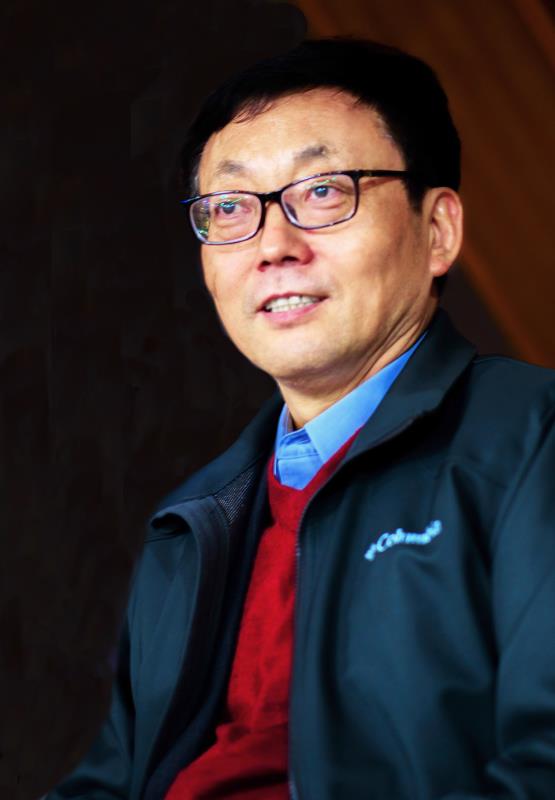}}]{Yu Liu} is Distinguished Professor and Dean of School of Kinesiology, Shanghai University of Sport, China. His research focuses on biomechanics of injuries, footwear biomechanics, neuromuscular control of human movement etc. He has published over 200 peer reviewed articles or book chapters, given over 130 international and domestic lectures. He is the Deputy Editors-in-chief of Journal of Sport and Health Science, and served on the editorial board of several national and international journals, including Footwear Science, Journal of Medical Biomechanics, China Sport Science, Chinese Journal of Sport Medicine etc. His research has been supported by large competitive grants from the Chinese Central Government, Shanghai Municipal Government and Nike Global Research Partner etc. He is Distinguished Professor of “Yangtze River Scholar” awarded by the China Ministry of Education, Outstanding teacher in Ten Thousand Talent Program, and owned Special government allowances of the State Council. During three decades of teaching and researching in the field of biomechanics, he served as Vice President of Asia Association of Coaching Science (AACS), Executive Council Member of International Society of Biomechanics (ISB), Vice President of the Chinese Society of Biomechanics in Sports, Member of the professional committee of biomechanics in the Chinese Society of Mechanics and Chinese Society of Biomedical Engineering.
\end{IEEEbiography}

\EOD

\end{document}